\documentclass[a4paper,fleqn]{article} 

\usepackage[utf8]{inputenc}
\usepackage{amsmath}
\usepackage{amsthm}
\usepackage{graphicx}
\usepackage{xcolor}
\usepackage{enumerate} 
\usepackage{todonotes}
\usepackage[nocompress]{cite}
\usepackage{fullpage}
\usepackage{hyperref}

\newcommand{\problem}{string indexing for top-$k$ close consecutive occurrences}
\newcommand{\shortproblem}{\textsc{Sitcco}}
\newcommand{\shortproblemfar}{\textsc{Sitfco}}

\newcommand{\str}[1]{\textrm{str}(#1)}
\newcommand{\loc}{\texttt{locus}}
\newcommand{\occ}{\textnormal{occ}}

\newcommand{\apex}{\textrm{apex}}
\newcommand{\offset}{\textrm{off}}
\newcommand{\ors}{\textsf{RangeSuccessor}}
\newcommand{\orp}{\textsf{RangePredecessor}}
 
\newtheorem{theorem}{Theorem}
\newtheorem{corollary}[theorem]{Corollary}
\newtheorem{lemma}[theorem]{Lemma}
\newtheorem{claim}[theorem]{Claim}
 
 \date{}

\title{String Indexing for Top-$k$ Close Consecutive Occurrences\footnote{An extended abstract appeared at FSTTCS 2020~\cite{bille2020topk}. The full version is published in Theor. Comput. Sci. (2022) \cite{DBLP:journals/tcs/BilleGPRS22}.}}

\author{
    Philip Bille\\\texttt{phbi@dtu.dk} \and
    Inge Li Gørtz\\\texttt{inge@dtu.dk} \and
    Max Pedersen\\\texttt{mhrpe@dtu.dk} \and
    Eva Rotenberg\\\texttt{erot@dtu.dk} \and
    Teresa Anna Steiner\\\texttt{terst@dtu.dk} 
}

\newcommand{\keywords}[1]
{
  \small	
  \textbf{\textit{Keywords---}} #1
}

\begin{document}
\setcounter{page}{0}
\maketitle

\begin{abstract}
The classic string indexing problem is to preprocess a string $S$ into a compact data structure that supports efficient subsequent pattern matching queries, that is, given a pattern string $P$, report all occurrences of $P$ within $S$. In this paper, we study a basic and natural extension of string indexing called the \problem\ problem (\shortproblem). Here, a consecutive occurrence is a pair $(i,j)$, $i < j$, such that $P$ occurs at positions $i$ and $j$ in $S$ and there is no occurrence of $P$ between $i$ and $j$, and their distance is defined as $j-i$. Given a pattern $P$ and a parameter $k$, the goal is to report the top-$k$ consecutive occurrences of $P$ in $S$ of minimal distance. The challenge is to compactly represent $S$ while supporting queries in time close to the length of $P$ and $k$. 
We give three time-space trade-offs for the problem. Let $n$ be the length of $S$, $m$ the length of $P$, and  $\epsilon\in(0,1]$. Our first result achieves $O(n\log n)$ space and optimal query time of $O(m+k)$.  Our second and third results achieve linear space and query times either $O(m+k^{1+\epsilon})$ or $O(m + k \log^{1+\epsilon} n)$. Along the way, we develop several techniques of independent interest, including a new  translation of the problem into a line segment intersection problem and a new recursive clustering technique for trees.  
\end{abstract}

\keywords{String indexing, pattern matching, consecutive occurrences}
\newpage
\section{Introduction}

The classic string indexing problem is to preprocess a string $S$ into a compact data structure that supports efficient subsequent pattern matching queries, that is, given a pattern string $P$, report all occurrences of $P$ within $S$. An \emph{occurrence} of $P$ within $S$ is an index $i$, $0\leq i < |S|$, such that $P=S[i\dots i+|P|-1]$.  In this paper, we introduce a basic extension of string indexing, where the goal is to report  \emph{consecutive occurrences} of the pattern $P$ that occur \emph{close} to each other in $S$. Here, a consecutive occurrence is a pair $(i,j)$, $i < j$, such that $P$ occurs at positions $i$ and $j$ in $S$ and there is no occurrence of $P$ between $i$ and $j$, and close to each other means that the distance $j-i$ between the occurrences should be small. More precisely, given a pattern $P$ and an integer parameter $k > 0$, define the \emph{top-$k$ close consecutive occurrences} of $P$ to be the $k$ consecutive occurrences of $P$ in $S$ with the smallest distances. Given a string $S$ the \emph{\problem\ } (\shortproblem) problem is to preprocess $S$ into a data structure that supports top-$k$ close consecutive occurrences queries. The goal is to obtain a compact data structure while supporting fast queries in terms of the length of the pattern $P$ and the number of reported occurrences $k$. For an example, see Figure~\ref{example}.

    \begin{figure}[h]
    \begin{align*}
    P&=\texttt{AN}\\
    S&=\underset{0}{\texttt{B}}\texttt{ATM{\color{red}A}}\underset{5}{\texttt{\color{red}N}}\texttt{ {\color{red}AN}D}\underset{10}{~}\texttt{{\color{red}AN}NA}\underset{15}{~}\texttt{SING}\underset{20}{~}\texttt{N{\color{red}ANA}}\underset{25}{\texttt{\color{red}N}}
    \texttt{{\color{red}AN}A }\underset{30}{\texttt{\color{red}A}}\texttt{{\color{red}N}D E}\underset{35}{\texttt{A}}\texttt{T B{\color{red}A}}\underset{40}{\texttt{\color{red}N}}\texttt{{\color{red}AN}AS}
    \end{align*}
    \caption{$P$ occurs at positions 4, 7, 11, 22, 24, 26, 30, 39 and 41 in $S$. The top $5$ close consecutive occurrences are $(22,24)$, $(24,26)$, $(39,41)$, $(4,7)$, and $(7,11)$, with the tie between $(7,11)$ and $(26,30)$ broken arbitrarily.}
    \label{example}
    \end{figure}

Surprisingly, the \shortproblem\ problem has not been studied before even though it is a natural variant  of string indexing and several closely related problems have been extensively studied (see related work below).

\subsection{Results and Techniques}
To state the complexity bounds, let $n$ and $m$ denote the lengths of $S$ and $P$, respectively. An immediate approach to solve the \shortproblem\ problem is to store the suffix tree of $S$ using $O(n)$ space. To answer a query on $P$ with parameter $k$, we traverse the suffix tree to find \emph{all} occurrences of $P$, construct the consecutive occurrences, and then sort these to output the top-$k$ close consecutive occurrences. Naively, this requires two sorts of size $\occ$, where $\occ$ is the total number of occurrences of $P$, giving a query time of $O(m+\occ\log\occ)$. Using more advanced data structures \cite{Brodal2009online,Select73}, the query time can be reduced to $O(m + \occ)$ while still using linear space.
Note that $\occ$ can be much larger than $k$. Alternatively, we can store at every node in the suffix tree the set of all consecutive occurrences sorted by distance using $O(n^2)$ space. To answer a query we find the node corresponding to $P$ and simply report the first $k$ of the stored consecutive occurrences in optimal $O(m + k)$ time. 

To achieve better trade-offs, one might try to use a strategy similar to range minimum query (RMQ), where the ranges are subsequent ranges in the suffix array and the values are distances between pairs of suffix indexes in $S$. However, there are several problems with that idea: first, there are $\Theta(|S|^2)$ possible pairs of suffix indexes within $S$, and it is not immediately clear how many of them can correspond to {\it consecutive} occurrences of a pattern (our arguments from Section \ref{sec:heavy-path} imply that this number is bounded by $O(n\log n)$). Secondly, when taking the union of two ranges, the set of closest (consecutive) pairs can change completely: consider for example the string $S=\texttt{A B A C A B A C D A B D A C D A B D A C}$. While the string \texttt{A} has occurrences $\{0,2,4,6,9,12,15,18\}$, the string \texttt{AB} has occurrences $\{0,4,9,15\}$ and \texttt{AC} has $\{2,6,12,18\}$. Note that for $P=\texttt{A}$, the top-3 consecutive occurrences are $(0,2)$, $(2,4)$ and $(4,6)$, while for \texttt{AB} they are $(0,4)$, $(4,9)$ and $(9,15)$ and for \texttt{AC} they are $(2,6)$, $(6,12)$ and $(12,18)$. Both the pairs and the distances are completely different between \texttt{A} and its extensions. Thus, there is an issue of non-decomposability, which is a main challenge in this particular problem. However, in the rest of our paper we will show that we can use suffix tree decompositions and amortized arguments to bound the number of changes that can happen in the set of consecutive occurrences of substrings corresponding to positions on some paths in the suffix tree.

We obtain the following significantly improved time-space trade-offs: 
\begin{theorem}
    \label{thm:main-result} Given a string $S$ of length $n$ and 
    $\epsilon$, $0<\epsilon\le 1$, 
    we can build a data structure that can answer 
    \emph{top-$k$ close consecutive occurrences queries} 
    using either
    \begin{enumerate}[(i)]
    \item $O(n\log n)$ space and $O(m+k)$ query time or 
    \label{thm:main-result-optimal-time}
    \item $O(\frac{n}{\epsilon})$ space and $O(m + k^{1 + \epsilon})$ query time.
    \label{thm:main-result-optimal-space}
    \item\label{thm:main-result-new-optimal-space} $O(\frac{n}{\epsilon})$ space and $O(m + k\log^{1+\epsilon}n)$ query time.
    \end{enumerate}
    Here, $m$ is the length of the query pattern.
    \end{theorem}
Hence, Theorem~\ref{thm:main-result}(\ref{thm:main-result-optimal-time}) achieves optimal query time using near-linear space. Alternatively,  Theorem~\ref{thm:main-result}(\ref{thm:main-result-optimal-space}) and (\ref{thm:main-result-new-optimal-space}) achieve linear space, for constant $\epsilon$, while supporting queries in near-optimal $O(m + k^{1 + \epsilon})$ and $O(m + k\log^{1+\epsilon}n)$ time, respectively. 

 To achieve Theorem~\ref{thm:main-result} we  develop several data structural techniques that may be of independent interest. First, we translate the problem into a line segment intersection problem on the heavy path decomposition of the suffix tree. This leads to the $O(n\log n)$ space and optimal query time bound of  Theorem~\ref{thm:main-result}(\ref{thm:main-result-optimal-time}). We note that Navarro~and~Thankachan~\cite{navarro2015reporting} used similar techniques for a  closely related problem (see related work below). To reduce space, we introduce a novel recursive   clustering method on trees. The decomposition partitions the tree into a hierarchy of depth $O(\log \log n)$ consisting of subtrees of doubly exponentially decreasing sizes. We show how to combine the  decomposition with the techniques of the simple algorithm from  Theorem~\ref{thm:main-result}(\ref{thm:main-result-optimal-time}) to obtain an $O(n \log \log n)$ space and $O(m + k^{1+\epsilon})$ query time solution. Then, we show how to efficiently compress the hierarchy of data structures into rank space leading to the linear space and $O(m+k^{1+\epsilon})$ query time bound of Theorem~\ref{thm:main-result}(\ref{thm:main-result-optimal-space}). Finally, we show how to use $O(\log n)$ cluster decompositions of varying parameters together with an orthogonal range successor data structure to obtain the $O(m+k\log^{1+\epsilon})$ time bound of Theorem~\ref{thm:main-result}(\ref{thm:main-result-new-optimal-space}).

We apply these techniques to three related problems: Firstly, we address the natural ``opposite'' problem of reporting the $k$ consecutive occurrences of \emph{largest} distance, which can be solved using similar but not identical techniques. Secondly, we apply our framework to the related problem of reporting consecutive occurrences with distances within a specified interval, considered by Navarro~and~Thankachan~\cite{navarro2015reporting}, and give an improvement for a special case. Finally, we show how this allows us to efficiently report all non-overlapping consecutive occurrences of a pattern.

\subsection{Related Work}
To the best of our knowledge, the \shortproblem\ problem has not been studied before, even though distances between occurrences is a natural extension for string indexing and several related problems have been studied extensively.

A closely related problem was considered by Navarro~and~Thankachan~\cite{navarro2015reporting}, who showed how to efficiently report consecutive occurrences with distances within a specified interval. They gave an $O(n\log n)$ space and $O(m + \occ)$ time solution, where $\occ$ denotes the number of reported consecutive occurrences. We note that their result can be adapted to the \shortproblem\ problem to achieve the same bounds as in  Theorem~\ref{thm:main-result}(\ref{thm:main-result-optimal-time}). However, our solution is simpler and does not rely on heavy word RAM techniques such as persistent van Emde Boas trees~\cite{Chan2013}. Our techniques can also be used to solve the problem considered by Navarro~and~Thankachan getting the same space and time bounds as they obtain, and we can achieve improved bounds in a special case (see Section~\ref{sec:extend}).

A lot of work has been done on the related problem of string indexing for patterns under various \emph{distance  constraints}, where the goal is to report occurrences of (one or more) patterns that are within a given distance or interval of distances of each other~\cite{bille2014substring,iliopoulos2009indexing,bader2016practical,caceres2020fast,bille2014string,lewenstein2011indexing, keller2007range}. An important difference between those works and our work is that all those solutions use time proportional to \emph{all} pairs of occurrences with distances in the given range, in contrast to only finding \emph{consecutive} occurrences. Note that if the goal is to find occurrences of a given maximal distance, one can find the close \emph{consecutive} occurrences first and then construct all pairs satisfying the constraint. 

Another line of related work is indexing collections of strings, called \emph{documents}. Here the goal is to find documents containing patterns subject to various constraints. For a comprehensive overview see the survey by Navarro~\cite{navarro2014spaces}. Several results on supporting efficient top-$k$ queries are known~\cite{munro2020ranked,shah2013top,hon2014space,biswas2018ranked,hon2010efficient,hon2014space,navarro2017time,Tsur13,Hon2013indexes,Hon2013faster,NavarroT14,MunroNNST17}. In this context the goal is to efficiently report the $k$ documents of smallest weight. The weights can depend on the query and can be the distance between the closest pair of occurrences of a given pattern~\cite{hon2014space,navarro2017time,shah2013top,hon2014space,navarro2017time,MunroNNST17}. The problem can be solved in linear space and optimal $O(k)$ time, in addition to finding the locus of the pattern in the suffix tree~\cite{shah2013top}. While this problem statement resembles ours, there is no direct translation from those results to our problem, since the documents are considered individually, and for a single document only the pair of occurrences with minimum distance within the document is considered.

Finally, we note that since the initial publication of the results in this paper, a subset of the authors have recently considered indexing for consecutive occurrences of two different patterns $P_1$ and $P_2$~\cite{BGPS2021}.

\subsection{Outline}
The paper is organized as follows. In Section~\ref{sec:prelims} we introduce some notation and recall results on string indexing. In Section~\ref{sec:heavy-path} we build a simple data structure and prove Theorem~\ref{thm:main-result}(\ref{thm:main-result-optimal-time}). In Section \ref{sec:fixed_k} we recall a method for tree clustering and show how to use it to solve a simplified version of the problem. In Section~\ref{sec:k_at_query} we introduce a recursive clustering method that allows us to use the ideas from Section~\ref{sec:fixed_k} on the actual problem. This gives an $O(n\log \log n)$ space and $O(m+k^2)$ time data structure. In Section~\ref{sec:saving_space}, we show how to reduce the space to linear while achieving the same query time, and then generalize the recursion to get Theorem~\ref{thm:main-result}(\ref{thm:main-result-optimal-space}) for any $0<\epsilon\leq 1$. In Section~\ref{sec:diff-tradeoff}  we give the linear space solution with $O(m + k \log^{1+\epsilon} n)$ query time. Finally, in Section \ref{sec:extend} we apply our techniques to related problems.

\section{Preliminaries}\label{sec:prelims}
We introduce some notation and  recall basic results from string indexing. 

A \emph{string} $S$ of length $n$ is a sequence $S[0]S[1]\dots S[n-1]$ of characters from an alphabet $\Sigma$. A contiguous subsequence $S[i,j]=S[i]S[i+1]\dots S[j-1]$ is a \emph{substring} of $S$. The substrings of the form $S[i,n]$ are the \emph{suffixes} of $S$.
 
The \emph{suffix tree}~\cite{weiner1973linear} is a compact trie of all suffixes of $S\$$, where \$ is a symbol not in the alphabet, and is lexicographically smaller than any letter in the alphabet. Using perfect hashing~\cite{FKS1984}, it can be stored in $O(n)$ space and solve the string indexing problem (i.e., find and report all occurrences of a pattern $P$) in $O(m+\occ)$ time, where $m$ is the length of $P$ and $\occ$ is the number of times $P$ occurs in $S$.
The \emph{suffix array} stores the suffix indices of $S\$$ in lexicographic order. The suffix tree has the property that the leaves below any node represent suffixes that appear in consecutive order in the suffix array.
Brodal~et~al.~\cite{Brodal2009online} show that there is a linear space data structure that allows outputting all entries within a given range of an array in sorted order using time linear in size of the output. This data structure on the suffix array together with the suffix tree can output all occurrences of a pattern \emph{sorted by text order} in $O(n)$ space and $O(m+\occ)$ time. 

For any node $v$ in the suffix tree, we define $\str{v}$ to be the string found by concatenating all labels on the path from the root to $v$.
The \emph{locus} of a string $P$, denoted $\loc(P)$, is the minimum depth node $v$ such that $P$ is a prefix of $\str{v}$. 

\section{A Simple $O(n \log n)$ Space Solution}\label{sec:heavy-path}
In this section, we present a simple solution that solves the \shortproblem\ problem in  $O(n\log n)$ space and $O(m + k)$ query time. This solution will be a key component in our more advanced structures in the following sections. We note that the results by Navarro~and~Thankachan~\cite{navarro2015reporting} for the related problem of reporting consecutive occurrences with distances within a specified interval can be modified to achieve the same complexities. However, our solution is simpler and does not rely on heavy word RAM techniques such as persistent van Emde Boas trees~\cite{Chan2013}.

Let $D(v)$ denote the set of consecutive occurrences of $\str{v}$. Naively, if we store for each node $v$ the set $D(v)$ in sorted order, we can directly answer a query for the top-$k$ close consecutive occurrences of a pattern $P$ by reporting the $k$ smallest elements in $D(\loc(P))$. This solves the problem in $O(n^2)$ space and $O(m + k)$ query time. The main idea in our simple solution is to build a heavy path decomposition of the suffix tree and compactly represent sets on the same path via a reduction to the orthogonal line segment intersection problem while maintaining optimal time queries. This is similar to the data structure by Navarro~and~Thankachan~\cite{navarro2015reporting}, but our reduction is different.

\paragraph*{Heavy path decomposition}
A \emph{heavy path decomposition} of a tree $T$ is defined as follows: Starting from the root, at every node, we choose the edge to the child with the largest subtree as \emph{heavy edge}, until we reach a leaf. Ties are broken arbitrarily. This defines a heavy path, and all edges hanging off the heavy path are \emph{light edges}. The root of a heavy path $h$ is called the \emph{apex} of the path, denoted $\apex(h)$. We then recursively decompose all subtrees hanging off the path. The heavy path decomposition has the following property:
    \begin{lemma}[Sleator and Tarjan~\cite{SLEATOR1983362}] Given a tree $T$ of size $n$ and a heavy path decomposition of $T$, any root-to-leaf path in $T$ contains at most $O(\log n)$ light edges.
    \end{lemma}

\paragraph*{Orthogonal line segment intersection} 
    Similarly as Navarro~and~Thankachan~\cite{navarro2015reporting}, we are going to reduce the problem to a geometric problem on orthogonal line segment intersection. Specifically, we are going to reduce to the following problem: Let $L$ be a set of $n$ vertical line segments in a plane with non-negative $x$-coordinates. The \emph{orthogonal line segment intersection problem} is to preprocess $L$ to support the query:
    \begin{itemize}
        \item \textsf{smallest-segments}($y_0,k$): return the first $k$ segments intersecting the horizontal line with $y$-coordinate $y_0$ in left-to-right order.
    \end{itemize}
    We will assume that $y_0$ is an integer, which suffices for our purpose. 
    Let $N$ be the maximum $y$-coordinate of a segment in $L$. 
    The following lemma follows easily from the results on partially persistent data structures by Driscoll~et~al.~\cite{DSST89}.
    
    \begin{lemma} \label{lem:line_segments} We can solve the line segment intersection problem as described above in $O(n+N)$ space and $O(k)$ time.
    
    \end{lemma}
    \begin{proof}
    Consider the $x$-coordinates as the elements of a set $X$ and the $y$-coordinate as time. The version of $X$ at a time $y_0$ contains exactly the $x$-coordinates of the line segments which intersect the horizontal line at $y_0$. 
    Now, the data structure is a partially persistent sorted doubly linked list $L$ on the elements of $X$. The elements are sorted in increasing order. Since we have at most $n$ line segments, the maximum size of $X$ as well as the maximum number of updates is $n$. Each update changes only $O(1)$ pointers in the linked list. Using the node copying technique from Driscoll~et~al.~\cite{DSST89} we can build a partially persistent linked list using $O(n)$ space. 
    To be able to find version $y_0$ in constant time, we keep an array of size $N$ with a pointer to the root of the version at each possible time step.  For a query $(y_0,k)$, use the sorted linked list $L$ to report the $k$ smallest elements at time $y_0$.
    
    If we use a linear scan to find the place to insert an element or find the element to be deleted we get a preprocessing time of $O(n^2)$. This can be improved to $O(n\log n)$ by using a (non-persistent)  balanced binary search tree  during the preprocessing holding all elements in the current version of $L$ together with a pointer to their node in the current version. When performing an update the binary search tree is used to find the position where the element must be inserted/deleted in $O(\log n)$ time. After the preprocessing step the tree is discarded.
    \end{proof}

\subsection{Data Structure}

We construct a heavy path decomposition of the suffix tree  $T$ of $S$.
Our data structure consists of a line segment data structure from Lemma \ref{lem:line_segments} for each heavy path of $T$ that compactly encodes the sets $D(v)$ for each node $v$ on the path.   

We describe the contents of the data structure for a single heavy path $h=v_1,\ldots,v_\ell$, where $v_1$ is the apex of the path. Consider a consecutive occurrence $(i,j)$ on some node on $h$ and imagine moving down the heavy path from top to bottom.
Either $(i,j)$ is a consecutive occurrence at the apex of $h$ or it will become a consecutive occurrence as soon as every suffix starting at an index between $i$ and $j$ has branched off the heavy path. Then it will stay a consecutive occurrence until either the suffix corresponding to $i$ or the suffix corresponding to $j$ (or both) branch off $h$. Thus, there exists an interval $[d_1,d_2]$ of depths on the heavy path such that $(i,j)\in D(v_d)$ if and only if $d\in [d_1,d_2]$. We say that $(i,j)$ is \emph{alive} in this interval. 

\begin{figure}[t]
        \centering
        \includegraphics[width=0.8\linewidth
        ]{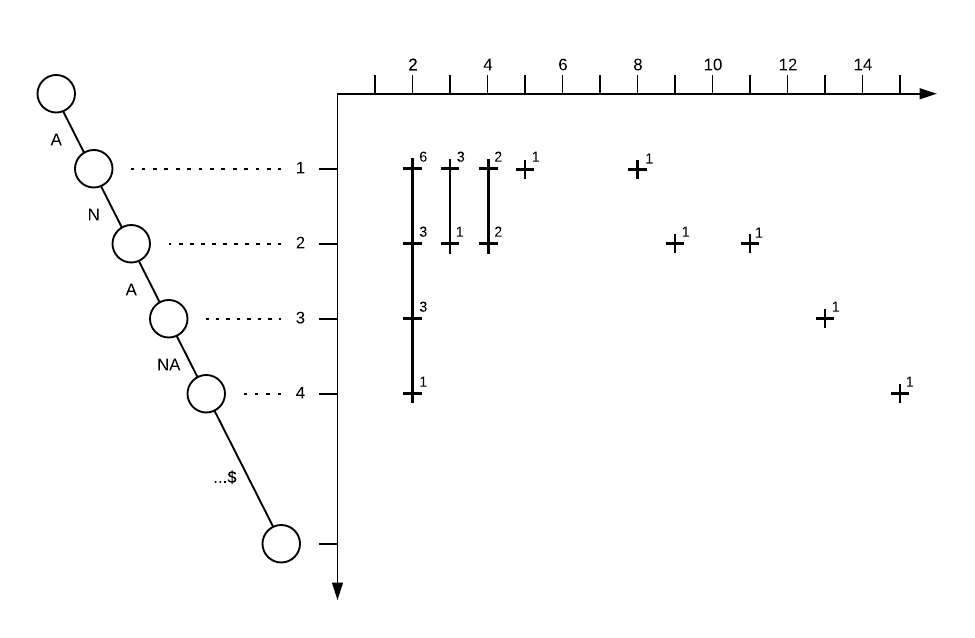}
        \caption{Line segments for a heavy path from the suffix tree for "\texttt{BATMAN-AND-ANNA-SING-NANANANA-AND-EAT-BANANAS}". Here, if we have overlapping line segments, we denote by a number how many consecutive occurrences the current segment corresponds to. At depth 1, we have a line segment corresponding to pairs of consecutive occurrences of string \texttt{A} - there are six pairs that have a distance of 2, three pairs that have a distance of 3, two pairs that have a distance of 4, and so on. At depth 2, we encode the consecutive occurrences of string \texttt{AN}. Some of them are the same as for string \texttt{A}.}
        \label{fig:path_solution}
    \end{figure}

We encode the consecutive occurrences by line segments in the plane which describe their distance and the interval in which they are alive along the heavy path. Conceptually, the $x$-coordinate in our coordinate system corresponds to the distance of a consecutive pair, and the $y$-coordinate corresponds to the depth on the heavy path. Now, for each consecutive occurrence $(i,j)$,  
we define a vertical line segment with $x$-coordinate set to its distance, and $y$-coordinate spanning the interval $[d_1,d_2]$, where $[d_1,d_2]$ is the interval  in which $(i,j)$ is alive. For an example, see Figure~\ref{fig:path_solution}. 
Our data structure for $h$ stores the above line segments in the line segment data structure from Lemma~\ref{lem:line_segments}. For each line segment in the data structure we store a pointer to the pair of occurrences it represents. The full data structure for $T$ consists of the line segment data structures for all of the heavy paths in $T$.

    \paragraph*{Space analysis}
    For a given heavy path $h$,  a leaf in the subtree of $\apex(h)$ can be in at most two consecutive occurrences in $D(\apex(h))$.  Consider a light edge $(v_d,u)$ leaving $h$ at depth $d$. Any leaf in the subtree rooted at $u$ can be part of at most two consecutive occurrences in $D(v_d)$. A single leaf can thus make at most two consecutive occurrences from $D(v_d)$ disappear in $D(v_{d+1})$ and at most one new consecutive occurrence appear. If we consider all leaves that leave $h$, 
    we therefore get at most three changes per leaf. Thus, for a given heavy path $h$ a leaf in the subtree of $\apex(h)$ can be in at most two consecutive occurrences in $D(\apex(h))$ and can cause at most three changes of line segments in the line segment data structure for $h$. 
    Since any root-to-leaf path can intersect at most $\log n$ heavy paths, any leaf can contribute $O(\log n)$ line segments.
    Overall, this means that there are at most $O(n\log n)$ line segments in total. 
    For a single heavy path $h$ the line segment data structure from Lemma~\ref{lem:line_segments} uses linear space in the number of segments and the length of $h$. The sum of the lengths of the heavy paths is $O(n)$, since the heavy paths are disjoint. Thus the total space usage 
    is $O(n\log n)$.    
    
    \subsection{Algorithm}
    Given a pattern $P$ and an integer $k$ we can now answer a query as follows. 
    We begin by finding $\loc(P)$ in the suffix tree. Let $h$ be the heavy path that the locus is on and let $d_P$ be the depth of $\loc(P)$ on $h$. We do a \textsf{smallest-segments}($d_P,k$) query on the line segment data structure stored for $h$ 
    and report the consecutive occurrences corresponding to the returned 
    line segments.

    \paragraph*{Correctness}
    By definition, $D(\loc(P))$ contains the consecutive occurrences of $P$. Thus, every consecutive occurrence of $P$ defines a line segment in the  data structure for $h$ and the horizontal line with $y$-coordinate set to $d_P$ intersects exactly those line segments. Since we set the $x$-coordinate of every line segment to the distance of its consecutive occurrence, the line segments are sorted left-to-right by increasing distance. Thus, the first $k$ line segments intersecting the horizontal line at $y=d_P$ correspond to the top-$k$ close consecutive occurrences.
         

    \paragraph*{Time analysis} 
The time for finding $\loc(P)$ in the suffix tree is $O(m)$. The time for querying the line segment data structure from Lemma~\ref{lem:line_segments} is $O(k)$, so the total time complexity is $O(m+k)$. This proves Theorem \ref{thm:main-result}(\ref{thm:main-result-optimal-time}).

\section{A Linear Space Solution for Fixed $k$}\label{sec:fixed_k}
In this section, we present a linear space and $O(m+k)$ time solution for the simpler problem where $k$ is known at construction time. 
That is, given a string $S$ and a positive integer $k$, we preprocess $S$ into a compact data structure such that given a pattern string $P$, we can efficiently find the top-$k$ close consecutive occurrences of $P$ in $S$.
This data structure demonstrates one of the key ideas that our final result builds on.

The main idea behind the data structure is to store the line segment solution from Section~\ref{sec:heavy-path} for some path segments of the suffix tree, such that all nodes that are not on these paths are within small subtrees.
For nodes within such small subtrees we can find all consecutive occurrences without spending too much time.
Specifically, we will partition the suffix tree into \emph{clusters}, satisfying some properties. We are going to define this cluster partition next.

\subsection{Cluster Partition}
For a connected subgraph $C\subseteq T$, a \emph{boundary node} $v$ is a node $v\in C$ such that either $v$ is the root of $T$, or $v$ has an edge leaving $C$ -- that is, there exists an edge $(v,u)$ in the tree $T$ such that $u\in T \setminus C$. 
A \emph{cluster} is a connected subgraph $C$ of $T$ with at most two boundary nodes.
A cluster with one boundary node is called a \emph{leaf cluster}. A cluster with two boundary nodes is called a \emph{path cluster}.
For a path cluster $C$, the two boundary nodes are connected by a unique path. We call this path the \emph{spine} of $C$. 
A \emph{cluster partition} is a partition of $T$ into clusters, i.e. a set $CP$ of clusters such that $\bigcup_{C\in CP}V(C)=V(T)$ and $\bigcup_{C\in CP}E(C)=E(T)$ and no two clusters in $CP$ share any edges. Here, $E(G)$ and $V(G)$ denote the edge and vertex set of a (sub)graph $G$, respectively. 
 We need the next lemma which follows from well-known tree decompositions  ~\cite{alstrup1997minimizing, alstrup2000maintaining, alstrup2002improved, frederickson1997ambivalent} (see Bille and Gørtz~\cite{bille2011tree}~Lemma~5.1 for a direct proof).

\begin{lemma}\label{lem:cluster}
    Given a tree $T$ with $n$ nodes and a parameter $\tau$, there exists a cluster partition $CP$ such that $|CP|=O(n/\tau)$ and every $C\in CP$ has at most $\tau$ nodes. Furthermore, such a partition can be computed in $O(n)$ time.
\end{lemma}

\subsection{Data Structure} 
\begin{figure}[t]
    \centering
    \includegraphics[width=0.5\textwidth]{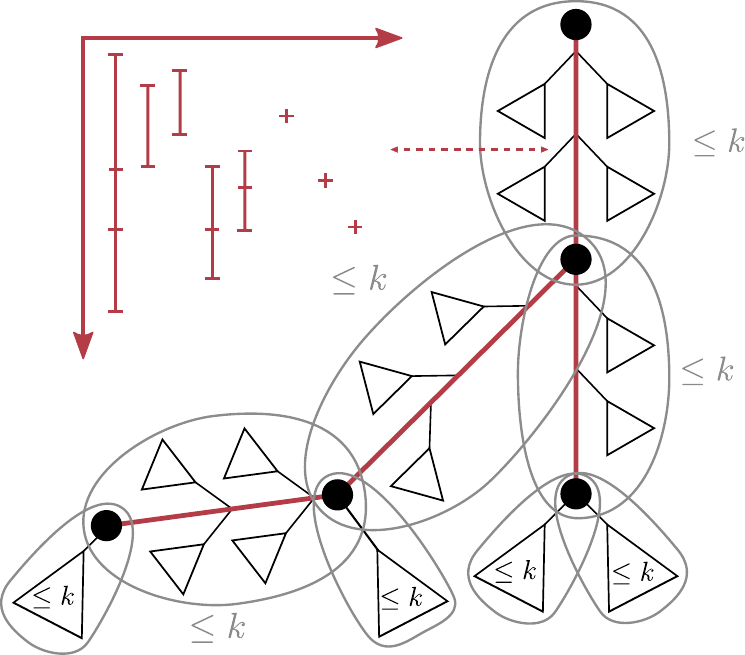}
    \caption{The suffix tree is divided into clusters (grey loops) of size $\le k$ which are either leaf clusters, or path clusters with spines marked in red. For every spine we store a line segment data structure, also marked in red.}
    \label{fig:cluster_1}
\end{figure}

For the suffix tree of $S$, we build a clustering as in Lemma~\ref{lem:cluster} with parameter $\tau$ set to $k$ to get $O(n/k)$ clusters of size at most $k$.
For the spine of every path cluster, we build a line segment data structure similar to the one from Section~\ref{sec:heavy-path}. The difference is that for any depth, we only maintain the line segments that correspond to the top-$k$ close consecutive occurrences for that depth.
Let $v_1,\dots, v_l$ denote the nodes on the spine, starting at the top boundary node. Note that for any consecutive occurrence that appears for the first time in $D(v_{d+1})$ there is a consecutive occurrence in $D(v_d)$ of smaller distance which is no longer present in $D(v_{d+1})$. 
It follows that, when moving down the spine, once a consecutive occurrence $(i,j)$ is amongst the $k$ closest, it will stay amongst the $k$ closest until suffix $i$ or $j$ branches off the spine. Thus, there exists an interval $[d_1,d_2]$ of consecutive depths such that $(i,j)$ is amongst the $k$ closest pairs in $D(v_d)$ if and only if $d\in [d_1,d_2]$. For a consecutive occurrence $(i,j)$ that is amongst the $k$ closest for any $v$ on the spine, we define a line segment where the $x$-coordinate is its distance and the $y$-coordinate is spanning the interval $[d_1,d_2]$, where $[d_1,d_2]$ is the interval in which $(i,j)$ is amongst the $k$ closest pairs. For these line segments we store the data structure from Lemma \ref{lem:line_segments}. 
Again, for each line segment we store the pair of occurrences it represents. We store this data structure for the spine of each cluster and for every node that is on that spine we store a pointer to the data structure. For boundary nodes that are on multiple spines we store a pointer to any one of them.
See Figure~\ref{fig:cluster_1} for an illustration of this structure. 
Additionally we store the suffix array and the sorted range reporting data structure of Brodal~et~al.\cite{Brodal2009online} on the suffix array.

\paragraph*{Space analysis}
We show that for every path cluster there are $O(k)$ line segments:  
We still have the property that a line segment only ends if a corresponding leaf branches off the spine. In that case, it might be replaced either by a new consecutive occurrence or by a consecutive occurrence that was there before but was not amongst the $k$ closest. 
Note that at any node on the spine except the boundary nodes, any subtrees branching off the spine are fully contained within the cluster, and as such have total size at most $k$. Between the top boundary node and the next node on the spine, we have no bound as to how many leaves can branch off --- however, since we only store line segments corresponding to the top-$k$ consecutive occurrences, at most 
$k$ line segments can be replaced by $k$ other line segments. For the rest of the spine, at most $k$ leaves can branch off in total. 
Every leaf that branches off can cause at most two line segments to end and two new line segments to begin. As such there can be at most $O(k)$ line segments.
As the size of the line segment data structure is linear in the number of line segments and in the length of the spine, any line segment data structure of a path cluster uses $O(k)$ space. 
As both the sorted range reporting data structure and the suffix array have linear space complexity, the complete data structure occupies $O((n/k)k+n)=O(n)$ space.

\subsection{Algorithm}
Given a pattern $P$ we can now answer the top-$k$ query.
We begin by finding $\loc(P)$ in the suffix tree.
If the locus is on a spine, we query the line segment data structure for that spine.
Otherwise the locus is either in a subtree hanging off a spine or in a leaf cluster. In both cases, there are at most $k$ occurrences of our pattern $P$. We find all occurrences of $P$ in text order, using the sorted range reporting data structure.
This allows us to report the consecutive occurrences:
Let $i_1, ..., i_l$ denote the leaves in text order, then the consecutive occurrences are $(i_1, i_2), (i_2, i_3), ... (i_{l-1}, i_l)$.
Note that $l \leq k$, since the size of the subtree is at most $k$.
\paragraph*{Correctness}
By construction, for any depth on a spine, the top-$k$ close consecutive occurrences of the corresponding substring will have corresponding line segments present at that depth in the line segment data structure. 
If the locus is on a spine, then by the arguments in Section~\ref{sec:heavy-path}, the line segment data structure will report the top-$k$ close consecutive occurrences.
If the locus is not on a spine, then there are at most $k$ occurrences of $P$ in total, since any subtree hanging off a spine and any leaf cluster has at most $k$ leaves. Thus, by constructing and reporting all consecutive occurrences of $P$ we report the top-$k$ close consecutive occurrences.
\paragraph*{Time analysis}
We find the locus in $O(m)$ time.
If we land on a spine we report in $O(k)$ time.
Otherwise, we are in a subtree of size at most $O(k)$ and thus $P$ has at most $k$ occurrences. Using sorted range reporting we can find the occurrences in text order using $O(k)$ time.
The total time for a query is thus $O(m + k)$. 

We are going to use this data structure with different parameters in Section \ref{sec:k_at_query}. For a general parameter $\tau$, we have the following lemma:
\begin{lemma} \label{lem:fixed_tau}
    For any positive integer $\tau$, there exists a cluster partition of the suffix tree and a linear space data structure with the following properties:
    \begin{enumerate}
        \item For any $k\leq\tau$ and $P$ such that $\loc(P)$ is on the spine of a cluster, we can report the top-$k$ close consecutive occurrences in $O(m+k)$ time. \label{lem:fixed_tau:case1}
        \item \label{loc_off_spine} For any $P$ such that $\loc(P)$ is not on a spine, we can report the top-$k$ close consecutive occurrences in $O(m+\tau)$ time. \label{lem:fixed_tau:case2}
    \end{enumerate}
\end{lemma}
\begin{proof}
We build the data structure described in this section for parameter $\tau$ taking the role of $k$. In case \ref{lem:fixed_tau:case1}, we query the line segment data structure for the depth of $\loc(P)$ on the path and $k$. Since $k\leq \tau$ this will correctly output the top-$k$ close consecutive occurrences of $P$. In case \ref{lem:fixed_tau:case2}, we have shown that we can construct the top-$\tau$ close consecutive occurrences. Using the linear time selection algorithm by Blum~et~al.~\cite{Select73} we can find the top-$k$ of those: We use the algorithm to find the consecutive occurrence of $k^{\textnormal{th}}$ smallest distance $d$; then we traverse all the consecutive occurrences and output those of distance $\leq d$. If needed, we crop the output to report no more than $k$ consecutive pairs.
\end{proof}

\section{An $O(n\log\log n)$ Space Solution for General $k$}\label{sec:k_at_query}

We now show how to leverage the solution from Lemma \ref{lem:fixed_tau} to obtain a data structure that can answer queries for any $k$.
The idea is to recursively cluster the suffix tree, such that we always either land on a spine with a sufficient number of consecutive occurrences stored, or in a sufficiently small subtree. 

\subsection{Data Structure}
\begin{figure}[t]
    \centering
    \includegraphics[width=0.5\textwidth]{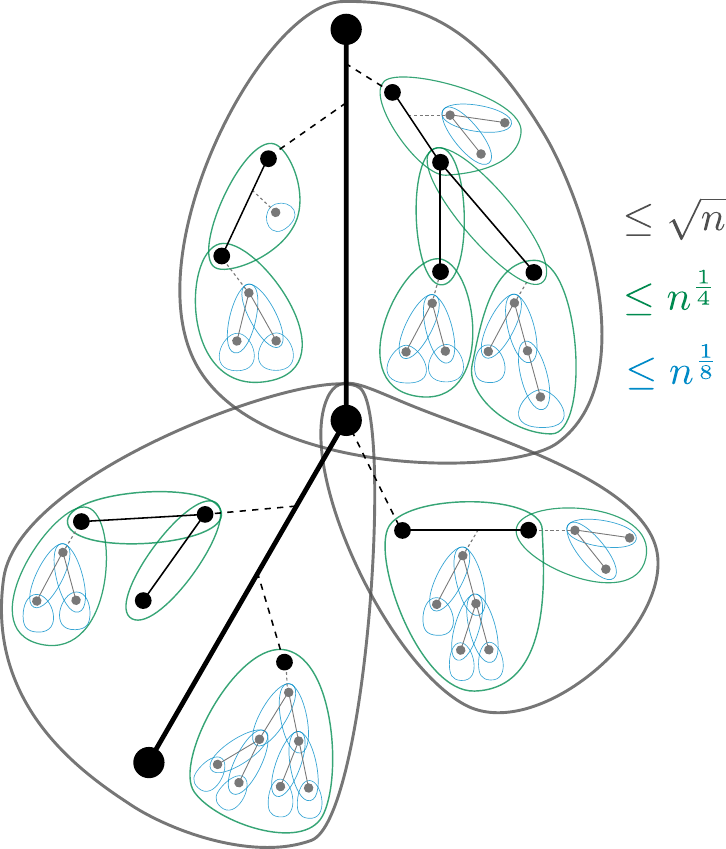}
    \caption{Here, we see the recursive clustering: The black clustering is the coarsest clustering and the green and blue are finer sub-clusterings.}
    \label{fig:rec_clusters}
\end{figure}

Our data structure consists of the suffix tree decomposed into clusters of decreasing size, with the line segment data structure stored for every spine as before. We build it in the following way.
First we build the solution from Lemma~\ref{lem:fixed_tau} with parameter $\tau_1 = \sqrt{n}$, resulting in clusters of size at most $\sqrt{n}$.
For every subtree hanging off a spine and every leaf cluster, we apply the solution with parameter $\tau_2 = \sqrt{\tau_1}$.
We keep recursively applying the solution with parameter $\tau_i = \sqrt{\tau_{i-1}}$ until reaching a constant cluster size.
For notational convenience, additionally define $\tau_0=n$.
See Figure~\ref{fig:rec_clusters} for an illustration of this data structure.
Again we additionally store the suffix array and the sorted range reporting data structure of Brodal~et~al.\cite{Brodal2009online} on the suffix array.
\paragraph*{Space analysis}
The suffix array and sorted range reporting structure occupy $O(n)$ space.
For a tree of size $\tilde{n}$ and any $\tau$, the data structure from Lemma~\ref{lem:fixed_tau} uses at most $O(\tilde{n})$ space.
Since at every recursion level, we build the data structure from Lemma~\ref{lem:fixed_tau} on non-overlapping subtrees of the suffix tree, every recursion level uses at most $O(n)$ space.
As the cluster size at every level of recursion is the square root of the previous cluster size, there are at most $O(\log \log n)$ levels.
The complete data structure thus uses $O(n \log \log n)$ space.

\subsection{Algorithm}
Given a query with pattern $P$ and parameter $k$, we can now answer in the following way.
As before, we begin by finding the locus of the pattern in the suffix tree.
This node is now either on the spine of some cluster or in a cluster of constant size.
If it is on the spine of a cluster of size $\tau_i$, and if $k \leq \tau_i$, then we query the line segment data structure for that spine, which allows us to report the top-$k$ close consecutive occurrences.
Otherwise, we find all occurrences of $P$ and construct the top-$k$ close consecutive occurrences by using linear time selection as in the proof of Lemma~\ref{lem:fixed_tau}.

\paragraph*{Correctness} The correctness of the algorithm follows by the same arguments as previous sections.

\paragraph*{Time analysis}
Finding the locus in the suffix tree takes $O(m)$ time. The locus is either on the spine of a cluster, or within a cluster of constant size. In a constant sized cluster, clearly we can do all operations described above in constant time.
If the locus is on the spine of a cluster with parameter $\tau_i$, and $k \leq \tau_i$, then we are in case 1 of Lemma~\ref{lem:fixed_tau} with $\tau=\tau_i$ and can report the top-$k$ close consecutive occurrences using a total of $O(m+k)$ time. 
If $k>\tau_i$, then we are in case 2 of Lemma~\ref{lem:fixed_tau} with $\tau=\tau_{i-1}$. Note that $\tau_{i-1}=\tau_i^2<k^2$. Therefore, we can find the top-$k$ close consecutive occurrences in $O(m+\tau_{i-1})=O(m+k^2)$ time.
In total, the worst case query time is then $O(m + k^2)$.
In summary, this gives the following result:
\begin{lemma}Given a string $S$ of length $n$, 
    we can build a data structure that can answer 
    \emph{top-$k$ close consecutive occurrences queries}
    using $O(n\log\log n)$ space and $O(m+k^2)$ query time. Here, $m$ is the length of the query pattern.
\end{lemma}
  
\section{A Linear Space Solution}\label{sec:saving_space}
We now show how to reduce the space consumption of the solution presented in Section \ref{sec:k_at_query}. Observe that in any cluster of level $i$, we only have $O(\tau_i)$ objects. If we can reduce all objects within a cluster to a ``universe size'' of $O(\tau_i)$ instead of $O(n)$, we can use $O(\tau_i\log \tau_i)$ bits instead of $O(\tau_i \log n)$ bits per cluster. In the following, consider a cluster $C$ of level $i$.

 \paragraph*{Reducing the line segment data structure}
    In the line segment data structure for cluster~$C$, by the analysis of previous sections, there are at most $O(\tau_i)$ line segments and $\tau_i$ different depths on the path. Let $c$ be a constant such that there are at most $c\tau_i$ line segments for each cluster. We map every unique $x$-coordinate of a line segment to a unique element in $\{1,\dots,c\tau_i\}$ in a way that preserves order. That is, map the minimum $x$-coordinate to 1, the smallest $x$-coordinate that is bigger than the minimum to 2, and so on. This gives us a modified line segment data structure that preserves the properties we need but is restricted to a $c\tau_i\times \tau_i$ grid.

\paragraph*{Reducing the leaf pointers}
    For any line segment, we have to store pointers that allow us to report the corresponding pair of consecutive occurrences. Doing so naively uses $2\log n$ bits per line segment. In the following, we show how to reduce that to $4 \log \tau_i$, for a cluster~$C$ of level $i$. The idea is to store the offset within the suffix array range defined by the top boundary node $r$ of $C$. More precisely, let $[a_r,b_r]$ be the range in the suffix array spanning the leaves below $r$. Then for any leaf $l$ in the subtree rooted at $r$ 
    define  $\offset(l)=SA^{-1}(l)-a_r$. By the way our recursion is defined, $C$ is fully contained in a subtree of size at most $\tau_i^2$, and thus $r$ has at most $\tau_i^2$ leaves below it. It follows that for any leaf $l$ in the subtree of $r$, $\offset(l)$ is a number between in $[0,\tau_i^2-1]$ and can be stored using $2\lceil\log \tau_i\rceil$ bits.
    
    \subsection{Data Structure}
    Our data structure is now defined as follows:  We have a clustering of the suffix tree as in Section \ref{sec:k_at_query}.
   For every spine on level $i$, we store the line segment data structure reduced to a $c\tau_i\times\tau_i$ grid. Every line segment corresponding to a pair $(i,j)$ stores the pair $(\offset(i),\offset(j))$ as additional information. 
   For every node on the spine, we store a pointer to the spine data structure and to the top boundary node of the spine. 
   Additionally, we store the suffix array and the sorted range reporting structure, as well as two integers for every node in the suffix tree, that define the range of leaves below the node in the suffix array.
   \paragraph*{Space analysis}
   The suffix array and the sorted range reporting data structure use space $O(n)$.
    Storing the range in the suffix array plus at most two pointers per node uses $O(n)$ space. 
    For a cluster $C$ of level $i$, we store the line segment data structure from Lemma~\ref{lem:line_segments} for a $c\tau_i\times \tau_i$ grid. Since the data structure from Lemma~\ref{lem:line_segments} works in the word RAM model (as do all data structures presented in this paper), we can store the data structure using $O(\tau_i \log \tau_i)$ bits. For each of the at most $c\tau_i$ line segments we store $4\log \tau_i$ bits for the encoding of the consecutive pair. Thus, we can store the data structure for cluster $C$ using $O(\tau_i \log \tau_i)$ bits.
    As in the previous section, at every recursion level, we cluster non-overlapping subtrees. The reduced cluster solution of a subtree of size $\tilde{n}$ with parameter $\tau_i$ uses $O(\frac{\tilde{n}}{\tau_i}\tau_i\log \tau_i)=O(\tilde{n}\log{\tau_i})$ bits. The total space for all clusters of level $i$ thus becomes $O(n\log \tau_i)$. Summing over all recursion levels, we get
    \[
    \sum_{i=0}^{\lfloor\log \log n\rfloor} O(n\log \tau_i) =
    \sum_{i=0}^{\lfloor\log \log n\rfloor} O\left(n\log n^{(1/2^i)}\right) =\sum_{i=0}^{\lfloor\log \log n\rfloor} \frac{1}{2^i} O(n\log n) = O(n\log n)\textnormal{ bits,}
   \]
    that is, $O(n)$ words.
    
   \subsection{Algorithm}
   We query the data structure as follows:
        If we land on a spine and $k\leq \tau_i$, we query the line segment data structure and get $k$ pairs of the form $(\offset(i),\offset(j))$.
        We then use the pointer to get to the root of the spine and use the range in the suffix array to translate each encoding back to the original suffix number, using constant time per leaf.
        Otherwise, we proceed as described in Section~\ref{sec:k_at_query}.
Since the decoding can be done in constant time per leaf, the time complexities are the same as in Section \ref{sec:k_at_query}.
    We have shown the following result:
    
\begin{lemma} Given a string $S$ of length $n$, 
    we can build a data structure that can answer 
    \emph{top-$k$ close consecutive occurrences queries}
    using  $O(n)$ space and $O(m+k^2)$ query time. Here, $m$ is the length of the query pattern.
\end{lemma}
In order to get Theorem \ref{thm:main-result}(\ref{thm:main-result-optimal-space}), we cluster according to a parameter $\epsilon$,  $0<\epsilon\leq 1$, using the following recursion:
\[
\hfill \tau_0=n \hfill\textrm{and}\hfill
 \tau_i=\tau_{i-1}^{\frac{1}{1+\epsilon}}\;.\hfill
\]
Hence, the total space in bits is now: 
%
\[\sum_{i=0}^{\lfloor\log_{1+\epsilon} \log n\rfloor} O \left(n\log n^{1/(1+\epsilon)^i}\right) =\sum_{i=0}^{\infty} \left(\frac{1}{1+\epsilon}\right)^i O(n\log n) = \left(1+\frac{1}{\epsilon}\right)O(n\log n),
\]
that is, $O\left(\frac{n}{\epsilon}\log n\right)$ bits, so $O\left(\frac{n}{\epsilon}\right)$ words.
For the query time, there are again two cases. In the case where $\loc(P)$ is on a spine with $k\leq \tau_i$, we get optimal $O(m+k)$ time, as before. For the other case, we have at most $\tau_{i-1}=\tau_i^{1+\epsilon}<k^{1+\epsilon}$ occurrences of $P$, which gives us a time complexity of $O(m+k^{1+\epsilon})$. 
This concludes the proof of Theorem \ref{thm:main-result}(\ref{thm:main-result-optimal-space}).

\section{A Different Tradeoff}\label{sec:diff-tradeoff}
In this section we give a solution query time $O(m + k \log^{1+\epsilon} n)$.
 The idea  is to store a finer set of cluster decompositions than in the previous section and store sublinear information for each cluster decomposition. Then we use a bounded number of orthogonal range successor queries in each cluster.

\paragraph{Orthogonal range successor}
The \emph{orthogonal range successor problem} is to  preprocess an array $A[0,\dots,n-1]$ into a data structure that  efficiently supports the following queries:
\begin{itemize}
    \item $\ors(a,b,x)$: return the successor of $x$ in $A[a,\dots, b]$, that is, the minimum $y> x$ such that there is an $i\in[a,b]$ with $A[i]=y$.
    \item $\orp(a,b,x)$: return the predecessor of $x$ in $A[a,\dots, b]$, that is, the maximum $y< x$ such that there is an $i\in[a,b]$ with $A[i]=y$.
\end{itemize}
Nekrich and Navarro~\cite{DBLP:conf/swat/NekrichN12} give a linear space data structure such that each range successor query takes $O(\log^{\epsilon} n)$ time.
We will use a range successor data structure on the suffix array to answer the following type of queries: Given an index $i$ and the suffix array range of a pattern $P$, find the next position in the text after $i$ where $P$ occurs.

\subsection{Data Structure}
We store the linear space range successor data structure from Nekrich and Navarro~\cite{DBLP:conf/swat/NekrichN12} and the sorted range reporting data structure by Brodal et al.~\cite{Brodal2009online} on the suffix array of $S$.
Further, we store the suffix tree of $S$ together with the following cluster decompositions. For each $\kappa=2,4,8,16,\dots,2^{\lfloor \log n\rfloor}$ we build the clustering decomposition of Lemma \ref{lem:cluster} of the suffix tree for cluster size $\tau=\kappa\log n$. For each boundary node $v$, we store the top-$\kappa$ close consecutive occurrences of $\str{v}$, sorted by text position. 
For each node $v$  in the suffix tree, we additionally store two bit vectors of length $\log n$. The first one is used to store for which $\kappa$ node $v$ is a boundary node, that is, 
the $i$th bit is set to 1 if $v$ is a boundary node in the cluster decomposition with $\kappa = 2^i$ and 0 otherwise. Similarly, 
the other bit vector stores for which $\kappa$ node $v$ is on a spine in the cluster decomposition with $\kappa = 2^i$.

\paragraph{Space analysis}
The suffix array, the range successor data structure and the sorted range reporting data structure all uses $O(n)$ space. The suffix tree together with the bit vectors saved in the nodes also uses $O(n)$ space. By Lemma \ref{lem:cluster}, there are $O(n/(\kappa\log n)$ boundary nodes in the cluster decomposition with cluster size $\kappa\log n$. For each boundary node we store $\kappa$ values and thus the space used for a fixed $\kappa$ is $O(\kappa n/(\kappa\log n))=O(n/\log n)$. 
There are $O(\log n)$ different values of $\kappa$ and therefore the total space is $O(n)$.

\subsection{Algorithm}
Given a pattern $P$ and parameter $k$ we can now answer the top-$k$ query in the following way.  As before, we begin by finding the locus of $P$ in the suffix tree. Then we find the smallest power of two bigger than $k$, i.e. $\kappa=2^{\lceil \log k\rceil}$, and consider the cluster decomposition defined for $\kappa$. 
There are two cases depending on whether $\loc(P)$ is on a spine in the cluster decomposition for $\kappa$ or not.

If $\loc(P)$ is not on a spine, then as in Lemma \ref{lem:fixed_tau} (\ref{lem:fixed_tau:case2}), there are at most $\tau=\kappa\log n$ leaves below $\loc(P)$, and we can construct the top-$k$ close occurrences in time $O(\kappa\log n)$. If $\loc(P)$ is on a spine, then we find the lower boundary node $b$ of the cluster (note we can do that by traversing the suffix tree and checking at most $O(\kappa\log n)$ nodes). 
We have the following property. 
\begin{claim} \label{claim} Each of the top-$k$ close consecutive occurrences of $P$ is either 
    \begin{enumerate}
    \item stored at $b$ or
    \item includes an occurrence of $P$ corresponding to a leaf within the cluster.
    \end{enumerate}
    \end{claim}
    \begin{proof} Any top-$k$ consecutive occurrence of $P$, where both occurrences are below $b$, is also among the top-$k$ consecutive occurrences of $\str{b}$. This is true because $P$ is a prefix of $\str{b}$, so the occurrences of $\str{b}$ is a subset of the occurrences of $P$. Thus any consecutive occurrence of $P$ where both occurrences are below $b$ is also a consecutive occurrence of $\str{b}$. A consecutive occurrences of $\str{b}$ that is not consecutive occurrences of $P$ must be split by an occurrence of $P$ that is not below $b$ giving rise to a least two closer consecutive occurrences of~$P$. Thus the $i$-closest occurrence of $\str{b}$ must have distance at least the same as the $i$-closest occurrence of~$P$.
     Every occurrence of $P$ that is not below $b$ is within the cluster.
    \end{proof}

We find all occurrences of $P$ that are within the cluster in text order using the sorted range reporting data structure. We can do this using two calls to the sorted range reporting data structure since the occurrences of $P$ within the cluster correspond to two intervals in the suffix array, namely the range of $\loc(P)$ minus the range of~$b$. For each such occurrence $i$, we use an orthogonal range successor query $j = \ors(\textrm{range}(\loc(P)),i) $ to find the next occurrence $j$ and then an orthogonal predecessor query $i' = \orp(\textrm{range}(\loc(P)),j) $ to find the last occurrence $i'$ before $j$. This gives us a consecutive occurrence $(i',j)$. To avoid recomputing the same consecutive occurrence we skip through the list until we get to an occurrence that is after $i'$.

Now we have the sorted list of the top-$\kappa$ consecutive occurrences of $\str{b}$ stored at $b$, and a sorted list of all consecutive occurrences of $P$ that include an occurrence corresponding to a leaf within the cluster. By Claim \ref{claim}, any  of the top-$k$ consecutive occurrences of $P$ is  part of one of these lists.
However, some of the consecutive occurrences of $\str{b}$ might not be consecutive occurrences of $P$, since $P$ might have extra occurrences inbetween. Let $(i,j)$ be a consecutive occurrence of $\str{b}$, and let $i'$ be the last occurrence of $P$ before $j$. If $i' \neq i$ then $i'$ is wihtin the cluster and thus $(i',j)$ is part of the consecutive occurrences we already computed. We merge the two lists, deleting all consecutive occurrences of $\str{b}$ that are not consecutive occurrences of $P$ and all duplicates.   We  then find the top-$k$ of remaining consecutive occurrences, by using the linear time selection algorithm  as in the proof of Lemma \ref{lem:fixed_tau}.

\paragraph{Time analysis} Finding the locus in the suffix tree takes $O(m)$ time. Finding $b$ and takes $O(\kappa\log n)$ time, and finding the occurrences of $P$ within the cluster using sorted range reporting also takes $O(\kappa\log n)$ time. We make $O(\kappa\log n) $ calls to the orthogonal range reporting data structure each using $O(\log^{\epsilon} n)$ time. In total the time for this is  $O(\kappa\log^{1+\epsilon} n)$. Merging the two lists and using the selection algorithm takes time linear in the total length of the two lists which is $O(\kappa\log n)$.   Since $\kappa < 2k$ the total time complexity $O(m + k\log^{1+\epsilon}n)$. This concludes the proof of  the main result.

\section{Extensions}\label{sec:extend}
Our results can be extended to a couple of related problems. In Section~\ref{sec:top-k_far}, we show how we can modify our data structure to solve the ``opposite'' problem of reporting the $k$ consecutive occurrences of \emph{largest} distance. The extension is quite natural, though it does require some careful analysis. In Section~\ref{sec:intervals}, we then relate the solutions from Section~\ref{sec:top-k_far} and the solutions to \shortproblem\ to the problem of finding consecutive occurrences with distances in a specified interval, considered by Navarro~and~Thankachan~\cite{navarro2015reporting}.
We show improved complexities for the special case where one of the interval bounds is known at indexing time. Finally, we show how to use those results to efficiently find all pairs of non-overlapping consecutive occurrences. 

\subsection{Top-$k$ Far Consecutive Occurrences}\label{sec:top-k_far}

Given a pattern $P$ and an integer parameter $k > 0$, define the \emph{top-$k$ far consecutive occurrences} of $P$ to be the $k$ consecutive occurrences of $P$ in $S$ with the largest distances. Given a string $S$ the \emph{string indexing for top-$k$ far consecutive occurrences problem} (\shortproblemfar) is to preprocess $S$ into a data structure that supports top-$k$ far consecutive occurrences queries. The goal is to obtain a compact data structure while supporting fast queries in terms of the length of the pattern $P$ and the number of reported occurrences~$k$. 
    
\paragraph*{Line segments and an $O(n\log n)$ space solution}
We can solve the \shortproblemfar\ problem using the same strategy as for the \shortproblem\ problem, with small modifications.
We need a similar data structure from Lemma~\ref{lem:line_segments} to report the line segments with largest $x$-coordinates. As previously, assume we are given a set $L$ of $n$ vertical line segments. We need a data structure for the following problem: 
\begin{itemize}
    \item \textsf{largest-segments}($y_0,k$): return the first $k$ segments intersecting the horizontal line with $y$-coordinate $y_0$ in right-to-left order.
\end{itemize}
As before, we can assume integer coordinates and let $N$ be the maximum $y$-coordinate of any line segment in $L$. 
\begin{lemma} \label{lem:line_segments_2} We can build a data structure that can answer largest-segment queries in $O(n+N)$ space and $O(k)$ time.
\end{lemma}
\begin{proof}
    We build the same data structure as in Lemma \ref{lem:line_segments}, but keep the partially persistent linked list sorted in decreasing order. The rest follows as before.
\end{proof}
Now, using this data structure in the solution described in Section \ref{sec:heavy-path}, we immediately get an analogous result for  \shortproblemfar :
    \begin{lemma}
        Given a string $S$ of length $n$, we can build a data structure that can answer \emph{top-$k$ far consecutive occurrences} queries using $O(n\log n)$ space and $O(m+k)$ query time.
    \end{lemma}

\begin{figure}[t]
    \centering
    \includegraphics[width=0.9\linewidth]{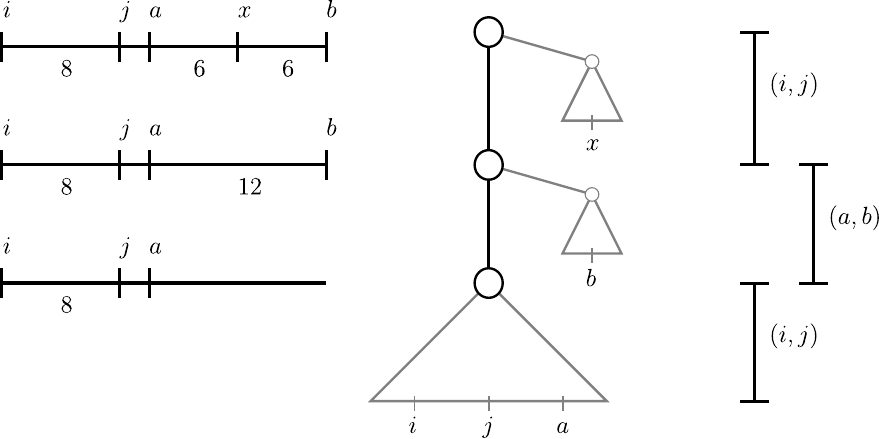}
    \caption{Illustration of a pair defining more than one line segment. To the left are the positions of the occurrences in $S$, in the middle is the spine of a cluster and to the right are the corresponding line segments. The pair $(i,j)$ is amongst the to $k$ farthest until the occurrence $x$ disappears, after which it is pushed out by the pair $(a,b)$. When $b$ then disappears, $(i,j)$ is again amongst the $k$ farthest.}
    \label{fig:recurring-pair-farthest}
\end{figure}

\paragraph*{Modifications to the linear space data structure}
Now we extend the cluster solutions from Sections \ref{sec:k_at_query} and \ref{sec:saving_space}.
We build the same recursive clusters as in Section \ref{sec:k_at_query}.
For each spine of a cluster of size $\tau_i$, we keep the line segments corresponding to the $\tau_i$ consecutive occurrences of largest distance at every depth on the spine.
That is, if a consecutive occurrence $(i,j)$ is among the $k$ farthest within $D(v_d)$ for some $v_d$ on the spine, define line segments for all maximal consecutive intervals $[d_1,d_2]$ such that $(i,j)$ is amongst the $k$ farthest within $D(v_d)$ for any $d\in [d_1,d_2]$. Again, the $x$-coordinate of the line segment is the distance $j-i$, and the $y$-coordinate spans $[d_1,d_2]$. Note that in this case, a consecutive occurrence might define more than one line segment.
See Figure~\ref{fig:recurring-pair-farthest} for an illustration of  pair defining more than one line segment.
We store these line segments in the data structure from Lemma~\ref{lem:line_segments_2}.

\paragraph*{Space analysis}
When moving down a spine from $v_{d-1}$ to $v_d$, only three different types of changes can happen to the set of the $k$ farthest consecutive occurrences. We again denote $D(v_d)$ to be the set of all consecutive occurrences of $\str{v_d}$. The possibles types of changes are then as follows.
\begin{itemize}
    \item A consecutive occurrence can be removed from the $k$ farthest because a consecutive occurrence of larger distance is added to $D(v_d)$. The consecutive occurrence of larger distance can only appear if an occurrence in between branched off. This leaf accounts for this change. A leaf can account for at most one such change, which triggers a line segment ending and a new line segment appearing at depth $d$.
    \item A consecutive occurrence $(i,j)$ can disappear because either $i$ or $j$ branched off. Then this leaf accounts for this change. A leaf can account for at most two such changes.
    \item A consecutive occurrence that was present in $D(v_{d-1})$ but not amongst the $k$ farthest  can be added to the $k$ farthest in $D(v_d)$. This can only happen if a consecutive occurrence of greater distance disappeared because one or both of its occurrences branched off.  Then this leaf accounts for this new line segment also, additional to the charge of the disappearing consecutive occurrence(s). A leaf can account for at most two such changes.
\end{itemize}
In total, any leaf can account for at most a constant number of changes.
Thus, we get the same space complexities as in Section \ref{sec:k_at_query}.
    
\paragraph*{Algorithm}
To answer a query we proceed as in Section \ref{sec:k_at_query}. 
We first find $\loc(P)$. 
If it is on a spine of a cluster of size $O(\tau_i)$ and $k<\tau_i$, we query the line segment data structure to report the top-$k$ far consecutive occurrences. Otherwise, we find all occurrences of $P$ in text order, construct the consecutive occurrences and use linear time selection to output the $k$ consecutive occurrences of largest distance. 
This is correct by the same arguments as Section \ref{sec:k_at_query}, and by similar arguments, achieves the same time complexities.
The rank space reduction from Section \ref{sec:saving_space} can be applied analogously. 
This gives us the following result:
\begin{theorem}
    Given a string $S$ of length $n$ and $\epsilon$, $0<\epsilon\le 1$, we can build a data structure that can answer \emph{top-$k$ far consecutive occurrences} queries using either
    \begin{enumerate}[(i)]
    \item $O(n\log n)$ space and $O(m+k)$ query time or 
    \item $O(\frac{n}{\epsilon})$ space and $O(m + k^{1 + \epsilon})$ query time.
    \end{enumerate}
    Here, $m$ is the length of the query pattern.
\end{theorem}

\paragraph{Remark} We note that the construction from Section~\ref{sec:diff-tradeoff} does not  generalize to the top-$k$ far consecutive occurrences problem, since the corresponding version of Claim~\ref{claim} does not hold. A top-$k$ far consecutive occurrence of $P$, where both occurrences are below $b$ is not necessarily among the top-$k$ far consecutive occurrences of $\str{b}$. A consecutive occurrence of $\str{b}$ can be split by an occurrence of $P$ not below $b$. This gives two smaller occurrences, and might cause the $(k+1)$th furthest occurrence below $b$ to be among the top-$k$ far occurrences of $P$. Each occurrence of $P$ from within the cluster can split a consecutive occurrence below $b$, and thus we would need to store $\Theta(\tau)$ occurrences below $b$, which no longer gives a linear space solution.

\subsection{Consecutive Occurrences with Gaps}\label{sec:intervals} 
Given a string $S$ the \textit{string indexing for consecutive occurrences with gaps problem} (\textsc{Sicog}) is to preprocess $S$ into a compact data structure, such that for any pattern $P$ and a range $[\alpha,\beta]$ we can efficiently find all consecutive occurrences of $P$ where the distance lies within $[\alpha,\beta]$.
The \textsc{Sicog} problem was considered by Navarro~and~Thankachan~\cite{navarro2015reporting} and they give an $O(n\log n)$ space and $O(m+\occ)$ time solution, where $\occ$ is the number of consecutive pairs with distance in $[\alpha, \beta]$. Using the data structure from Section \ref{sec:heavy-path}, we get an $O(n\log n)$ space and $O(m+\log n + \occ)$ time solution for the \textsc{Sicog} problem, which can be optimized using the same strategy as in \cite{navarro2015reporting} to achieve the same complexities. However, for a special case of the problem where either $\alpha$ or $\beta$ is known at indexing time we can get a similar trade-off as for the \shortproblem\ problem: 
We first the describe our solution for the fixed-$\alpha$ variant using the techniques from Sections \ref{sec:k_at_query} and \ref{sec:saving_space}, and then the fixed-$\beta$ variant follows by applying the same ideas combined with the data structure from Section~\ref{sec:top-k_far}.

\paragraph*{Data structure}
We build the same data structure
as in Section~\ref{sec:k_at_query}, with a slight modification. 
In the line segment data structure stored at every spine, instead of storing the $\tau_i$ closest pairs, we store the $\tau_i$ closest pairs that have distance $\geq \alpha$.
This clearly occupies no more space than the solution from Section~\ref{sec:k_at_query} and we can still apply the space optimizations of Section~\ref{sec:saving_space}.
    
\paragraph*{Algorithm}
Given $P$ and $\beta$, we can now answer a query as follows.
We begin by finding $\loc(P)$.
If it is in a subtree of constant size, we construct all the consecutive occurrences of $P$ and report those that have distance within $[\alpha,\beta]$.
If it is on a spine of a cluster of size $\tau_i$, we query the line segment data structure. For every consecutive occurrence we find, we check if the distance is $\leq \beta$.
If we encounter a pair with distance $>\beta$, we stop reporting. 
If all $\tau_i$ consecutive occurrences at $\loc(P)$ have distance $\leq \beta$, we then find all the the consecutive occurrences of $P$, just as in Section~\ref{sec:k_at_query}, and scan them once to report all the consecutive occurrences with distance in $[\alpha,\beta]$.  

For the analysis, define a \emph{relevant pair} to be a consecutive occurrence with a distance in $[\alpha,\beta]$.
If there are less than $\tau_i$ relevant pairs for any locus, then they will all be stored and reported by the line segment data structure.
As they are stored in order of increasing distance, once we reach a pair with distance $>\beta$, no further relevant pairs exist.
If there are more than $\tau_i$ relevant pairs, then we consider all occurrences of the pattern and report from those.
Thus we always answer the query correctly.

If there is a consecutive occurrence among the $\tau_i$ line segments with distance $>\beta$, we spend time $O(m+\occ)$ finding the locus and querying the line segment data structure.
Otherwise we have that $\occ\geq \tau_i$, and thus $\occ^{1+\epsilon}\geq \tau_{i-1}$. As before, $P$ has at most $\tau_{i-1}$ occurrences. Therefore, by the same arguments as in the previous section, we get the following result: 
\begin{theorem}
    Given a string $S$ of length $n$ and $\alpha>0$, we can build for any $\epsilon$ satisfying $0<\epsilon\leq 1$ an $O(\frac{n}{\epsilon})$ space data structure that can answer the following query in $O(m + \occ^{1 + \epsilon})$ time: For a query pattern $P$ and $\beta\geq\alpha$, report all consecutive occurrences of $P$ in $S$ where the distance lies in $[\alpha,\beta]$. Here, $m$ is the length of the pattern and $\occ$ is the number of reported occurrences.
\end{theorem}
By combining the same arguments with the solution for top-$k$ far consecutive occurrences, we get the following result for $\beta$ fixed at indexing time:
\begin{theorem}\label{cor:fixed_beta}
    Given a string $S$ of length $n$ and $\beta>0$, we can build for any $\epsilon$ satisfying $0<\epsilon\leq 1$ an $O(\frac{n}{\epsilon})$ space data structure that can answer the following query in $O(m + \occ^{1 + \epsilon})$ time: For a query pattern $P$ and $\alpha$ where $0<\alpha\le\beta$, report all consecutive occurrences of $P$ in $S$ where the distance lies in $[\alpha,\beta]$. Here, $m$ is the length of the pattern and $\occ$ is the number of reported occurrences.
\end{theorem}

We note that the construction from Section~\ref{sec:diff-tradeoff} does not  generalize to the problem where $\alpha$ is fixed, as the corresponding version of Claim~\ref{claim} do not hold in this case. A consecutive occurrence below $b$ of distance at least $\alpha$ can be split by an occurrence of $P$ from within the cluster introducing two new consecutive occurrences that might both have distance less than~$\alpha$. 

We can, however, get a solution to the problem when $\alpha = 1$. Using the data structure from Section~\ref{sec:diff-tradeoff}, we can get all consecutive occurrences that are at most $\beta$ apart as follows. We query the data structure for the top-$k$ close consecutive occurrences with $k=1, 2, 4, \ldots$, each time checking if all the top-$k$ close consecutive occurrences have distance at most $\beta$. As soon as we find a consecutive occurrence that has a distance more than $\beta$ among our top-$k$ close consecutive occurrences, we stop and report all the occurrences found in this last call that have distance at most $\beta$. This way we ensure that $k \leq 2\cdot \occ$. We only find the locus once. The total query time is $ O(m + \sum_{i=0}^{\lceil\log \occ \rceil}2^i\log^{1+\epsilon}n) = O(m + \occ\log^{1+\epsilon}n)$. 

\begin{lemma}
Given a string $S$ of length $n$ we can build an $O(n)$ space data structure that can answer the following query in $O(m + \occ\log^{1+\epsilon}n)$ time: For a query pattern $P$ and $\beta > 0$, report all consecutive occurrences with distance at most $\beta$.
\end{lemma}

\paragraph*{Non-overlapping consecutive occurrences}
A natural and well-studied variant of string indexing is the problem of finding sets of \emph{non-overlapping} occurrences of a pattern $P$. Here, a set of non-overlapping occurrences is a set of occurrences $\{i_1,\dots, i_k\}$ of $P$ such that the distance between any two of them is at least $|P|$. Several papers study the problem of finding the set of non-overlapping occurrences of maximum size~\cite{keller2007range, cohen2009range,GangulyST20,HooshmandAKT18}. Note that Theorem \ref{cor:fixed_beta} applied to $\alpha=|P|$ solves a different variant of finding sets of non-overlapping occurrences: Namely, finding all pairs of non-overlapping \emph{consecutive} occurrences. We call this problem the \emph{string indexing for non-overlapping consecutive occurrences problem} (\textsc{Sinoco})
. The \textsc{Sinoco} problem is inherently different from finding the maximum set of non-overlapping occurrences: For example, the maximum set of non-overlapping occurrences of the pattern $P=\texttt{NANA}$ in the string $S=\texttt{NANANANA}$ has size 2. However, there are no non-overlapping \emph{consecutive} occurrences. To the best of our knowledge, the \textsc{Sinoco} problem has not been studied before. An immediate corollary of the results in~Navarro~and~Thankachan~\cite{navarro2015reporting} and Theorem \ref{cor:fixed_beta} gives the following trade-offs for solving \textsc{Sinoco}:
\begin{corollary}\label{cor:non-overlap}
Given a string $S$ of length $n$ and $\epsilon$, $0<\epsilon\le 1$, we can build a data structure that can find all non-overlapping consecutive occurrences of a query pattern $P$ using either
    \begin{enumerate}[(i)]
\item $O(n\log n)$ space and $O(m+\occ)$ query time or 
\label{cor:non-overlap-optimal-time}
\item $O(\frac{n}{\epsilon})$ space and $O(m + \occ^{1 + \epsilon})$ query time.
\label{cor:non-overlap-optimal-space}
\end{enumerate}
Here, $m$ is the length of the query pattern and $\occ$ is the number of reported occurrences.
\end{corollary}
\begin{proof}
Apply the results in \cite{navarro2015reporting} and Theorem~\ref{cor:fixed_beta} with $\beta=n$ and $\alpha=|P|$.
\end{proof}
  
\section{Conclusion and Open Problems}
We have introduced 
the natural problem of \problem,\ 
and have given both a near-linear space solution achieving optimal query time and a linear space solution achieving a query time that is close to optimal. Using these techniques, we have given new solutions for the problem of string indexing for consecutive occurrences with gaps (\textsc{Sicog}). Furthermore, we have introduced the problem of finding all non-overlapping consecutive occurrences of a pattern (\textsc{Sinoco}) and showed that it can be reduced to a special case of \textsc{Sicog}. 

These results open interesting new directions for further research. The most obvious open problem is to see whether it is possible to further improve the results for the main problem considered in this paper, especially,  achieve linear space and optimal query time simultaneously. 
 Secondly, it is still open whether it is possible to get an $O(m+\occ)$ time and linear space solution for the special case of the \textsc{Sicog} problem where one of the interval endpoints is fixed, or even $o(n\log n)$ space for the general problem. For the \textsc{Sinoco} problem, one might find better solutions that do not reduce it to \textsc{Sicog} but use additional insights about the specific structure of the problem.

\section*{Acknowledgments}
We thank anonymous reviewers of an earlier draft of this paper for their insightful comments and suggestions for improvement. Philip Bille, Inge Li G{\o}rtz, Max Pedersen and Eva Rotenberg were partially supported by the Danish Research Council grant { \it Adaptive Compressed Computation} (DFF-8021-002498).

\bibliographystyle{plainurl}
\bibliography{References}

\end{document}